\documentclass[12pt]{iopart}
\usepackage{graphicx}
\usepackage{caption}
\usepackage[table]{xcolor}
\usepackage[colorinlistoftodos]{todonotes}
\usepackage{cite}

\begin{document}

\title[]{How high is a MoSe$_2$ monolayer?} 

\author{Rikke Plougmann$^1$ \footnote{These authors contributed equally to this work. \label{note1}} \footnote{Present address: Department of Physics, Technical University of Denmark, Fysikvej, Building 311, 2800 Kgs. Lyngby, Denmark.}, Megan Cowie$^1$ $\ddagger$, Yacine Benkirane$^1$, L\'eonard Schu\'e$^2$, Zeno Schumacher$^1$\footnote{Present address: Institute of Quantum Electronics, ETH Zürich, Auguste-Piccard-Hof 1, 
8093 Zürich, Switzerland.} and Peter Grütter$^1$}

\address{$^1$ Department of Physics, McGill University, 3600 Rue University, Montr\'eal, Qu\'ebec H3A 2T8, Canada}
\address{$^2$ D\'epartement de Chimie and Regroupement Qu\'eb\'ecois sur les Mat\'eriaux de Pointe (RQMP), Universit\'e de Montr\'eal, C.P. 6128, Succursale Centre-Ville, Montr\'eal, Qu\'ebec H3C 3J7, Canada}
\ead{rikplo@dtu.dk}

\begin{abstract}

Transition metal dichalcogenides (TMDCs) have attracted significant attention for optoelectronic, photovoltaic and photoelectrochemical applications. The properties of TMDCs are highly dependent on the number of stacked atomic layers, which is usually counted post-fabrication, using a combination of optical methods and atomic force microscopy (AFM) height measurements. Here, we use photoluminescence spectroscopy and three different AFM methods to demonstrate significant discrepancies in height measurements of exfoliated MoSe$_2$ flakes on SiO$_2$ depending on the method used. We highlight that overlooking effects from electrostatic forces, contaminants and surface binding can be misleading when measuring the height of a MoSe$_2$ flake. These factors must be taken into account as a part of the protocol for counting TMDC layers.

\end{abstract}

\submitto{\NT}
\maketitle


\section{Introduction}


Transition metal dichalcogenides (TMDCs) are of increasing interest as promising contenders for a wide range of optoelectronic and electrochemical applications due to their strong light-matter interactions, tunable optical bandgaps and flexible two-dimensional (2D) structure \cite{Li2018EngineeringConversion,Das2019ThePhotovoltaics}. Molybdenum diselenide (MoSe$_2$) is one of the most promising TMDC candidates for photoelectrochemical energy conversion due to its photocatalytic properties and the electrical conductivity of Se \cite{Eftekhari2017MolybdenumOptoelectronics,Li2018EngineeringConversion}. MoSe$_2$ monolayers have a hexagonal crystal structure that consists of top and bottom Se layers sandwiching a Mo layer \cite{Wang2012ElectronicsDichalcogenides}. Stacked MoSe$_2$ layers are weakly coupled via van der Waals coupling\cite{Le2016NonlinearMoSe2} and have an interlayer spacing in bulk of 0.65~nm\cite{James1963TheMoSe2,Tomm2005CrystalPhotovoltaics,Mahatha2012ElectronicStudies,Roy2016StructuralEpitaxy,Jiang2020Mini-review:Biosensors}. 

The number of stacked atomic layers in TMDC samples can drastically affect their properties \cite{Choi2017RecentApplications,Zhang2017OpticalSubstrates,Wang2012ElectronicsDichalcogenides}. For example, a MoSe$_2$ monolayer has a direct bandgap of 1.55~eV (800~nm) \cite{Tongay2012ThermallyMoS2}, which transitions to an indirect bandgap as the number of layers increases to 2 and 3 \cite{Tongay2012ThermallyMoS2,Sha2017Layer-by-layerEtching,Tonndorf2013PhotoluminescenceWSe_2,Le2016NonlinearMoSe2}. For both fundamental studies and device functionality, it is therefore vital to be able to confirm the number of layers present in TMDC samples. In addition, it is important to understand how the properties of the different layers in stacked TMDC structures could be influenced during fabrication and from interactions with the underlying substrate \cite{Pollmann2020ApparentExfoliation}. 

As with other 2D materials, TMDC layers can be synthesized both via chemical vapor deposition (CVD) \cite{Lu2014Large-areaSubstrates,Wang2019SynthesisDeposition,Wang2014ChemicalMoSe2} and, most commonly, using mechanical exfoliation \cite{Pollmann2020ApparentExfoliation,Choi2017RecentApplications,Li2018EngineeringConversion,Das2019ThePhotovoltaics,Ottaviano2017MechanicalRevisited,Larentis2012Field-effectLayers,Tongay2013DefectsExcitons,Sha2017Layer-by-layerEtching}, in which mono- or few-layer flakes are transferred onto a substrate, often SiO$_2$/Si. Studies of factors from the fabrication process influencing the measured step heights and optical properties of monolayers have been emerging mostly on MoS$_2$ \cite{Ottaviano2017MechanicalRevisited,Donnelly2020AchievingMoS2}, whereas the literature for other TMDCs such as MoSe$_2$ is limited in comparison. E. Pollman et al. \cite{Pollmann2020ApparentExfoliation} carried out a study on MoS$_2$ comparing the properties of samples fabricated from both techniques, and showed that a major difference is the presence of intercalated water on exfoliated monolayers, leading to both an increased step height and a decreased intensity in photoluminescence spectroscopy \cite{Pollmann2020ApparentExfoliation}. A further challenge is that samples are often handled in air, leading to airborn contaminants \cite{Purckhauer2019AnalysisConditions}. Interestingly, MoSe$_2$ has been shown to be even more susceptible to surface contaminants than sulfur-based TMDCs \cite{Purckhauer2019AnalysisConditions}, so understanding how contaminants and other effects may influence the height (and thus layer number) measurements for MoSe$_2$ is highly relevant.

\subsection{Methods for counting layers}

The number of layers present in TMDC samples is usually measured using a combination of optical microscopy \cite{Castellanos-Gomez2014DeterministicStamping,Nagler2017InterlayerHeterostructure}, photoluminescence spectroscopy (PL) and/or Raman spectroscopy together with step height measurements made by atomic force microscopy (AFM) \cite{Tonndorf2013PhotoluminescenceWSe_2,Arora2015ExcitonLimit,Lu2014Large-areaSubstrates,Wang2016}. In optical microscopy images, the contrast difference between layers can be analysed for the red, green and blue channels of the image \cite{Taghavi2019ThicknessMaterials,Zhang2017OpticalSubstrates,Wu2017RapidDichalcogenides}. AFM is used to measure the height of a flake, and comparing that to the expected interlayer spacing of the 2D material. There is much discrepancy among AFM height measurements of mechanically exfoliated TMDCs on SiO$_2$: Some show heights between 0.65~-~1.0~nm\cite{Tongay2013DefectsExcitons,Sha2017Layer-by-layerEtching,Arora2015ExcitonLimit} while others are much larger, ranging between 2~-~3~nm \cite{Ottaviano2017MechanicalRevisited,Li2012QuantitativeSubstrates}. Elevated heights are often attributed to surface contaminants \cite{Ottaviano2017MechanicalRevisited,Sha2017Layer-by-layerEtching,Li2018EngineeringConversion,Haigh2012Cross-sectionalSuperlattices}.

PL is another widely used technique used to characterize TMDC layer numbers. The position of the monolayer peak in a PL spectrum for exfoliated MoSe$_2$ samples has been observed at room temperature in the range 788~-~816~nm \cite{Sha2017Layer-by-layerEtching,Tonndorf2013PhotoluminescenceWSe_2,McCreary2018A-Monolayers,Kumar2014ExcitonMoSe2,Le2016NonlinearMoSe2,Silva2020MorphologicalHeterojunctions,Lu2014Large-areaSubstrates,Wang2019SynthesisDeposition}. The PL intensity can be a affected by the presence of defects \cite{Wang2014ChemicalMoSe2} and adsorbates \cite{Tonndorf2013PhotoluminescenceWSe_2} and is expected to decrease as the number of layers increases to 2 and 3 \cite{Tongay2012ThermallyMoS2,Sha2017Layer-by-layerEtching,Tonndorf2013PhotoluminescenceWSe_2,Le2016NonlinearMoSe2}. There is a lack of consensus in the literature regarding how to interpret PL spectra for 2, 3, and N layers of MoSe$_2$. For MoSe$_2$ flakes mechanically exfoliated onto SiO$_2$/SiO, P. Tonndorf et al. \cite{Tonndorf2013PhotoluminescenceWSe_2} show a 15~nm redshift of the dominant A peak for a bilayer and significant broadening and flattening of the A peak for 3 layers. In contrast, Y. Sha et al. \cite{Sha2017Layer-by-layerEtching} measured only a few nm redshift as the number of MoSe$_2$ layers increases. S. Tongay et al. \cite{Tongay2012ThermallyMoS2} also show a negligible shift in position of the peak as the number of MoSe$_2$ layers increases from 1 to 3 layers. In some cases, observations from bilayers have exhibited the properties of a monolayer, which is thought to be due to the formation of pockets between the layers during exfoliation\cite{Arora2015ExcitonLimit}. The degree of coupling between exfoliated MoSe$_2$ layers has been shown to significantly influence their photoluminescence, as demonstrated by S. Tongay et al. using thermally controlled coupling \cite{Tongay2012ThermallyMoS2}.

In this work, two MoSe$_2$ flakes were measured using optical microscopy, PL spectroscopy and three different modes of AFM: tapping mode, non-contact AFM (nc-AFM) and Kelvin probe force microscopy (KPFM). The presence of electrostatic forces can significantly influence the topographic height profile measured using nc-AFM \cite{Sadewasser2004InfluenceMicroscopy,SadewasserCorrectMicroscopy}. In KPFM, the net electrostatic force is minimized using a feedback loop to apply a DC bias, and KPFM thereby provides information about contact potential variations over the sample surface, either due to nonhomogeneities in the sample material or localized charging or polarization effects.

\section{Results \& Discussion}


\begin{figure}
\centering
\includegraphics[width=1\textwidth]{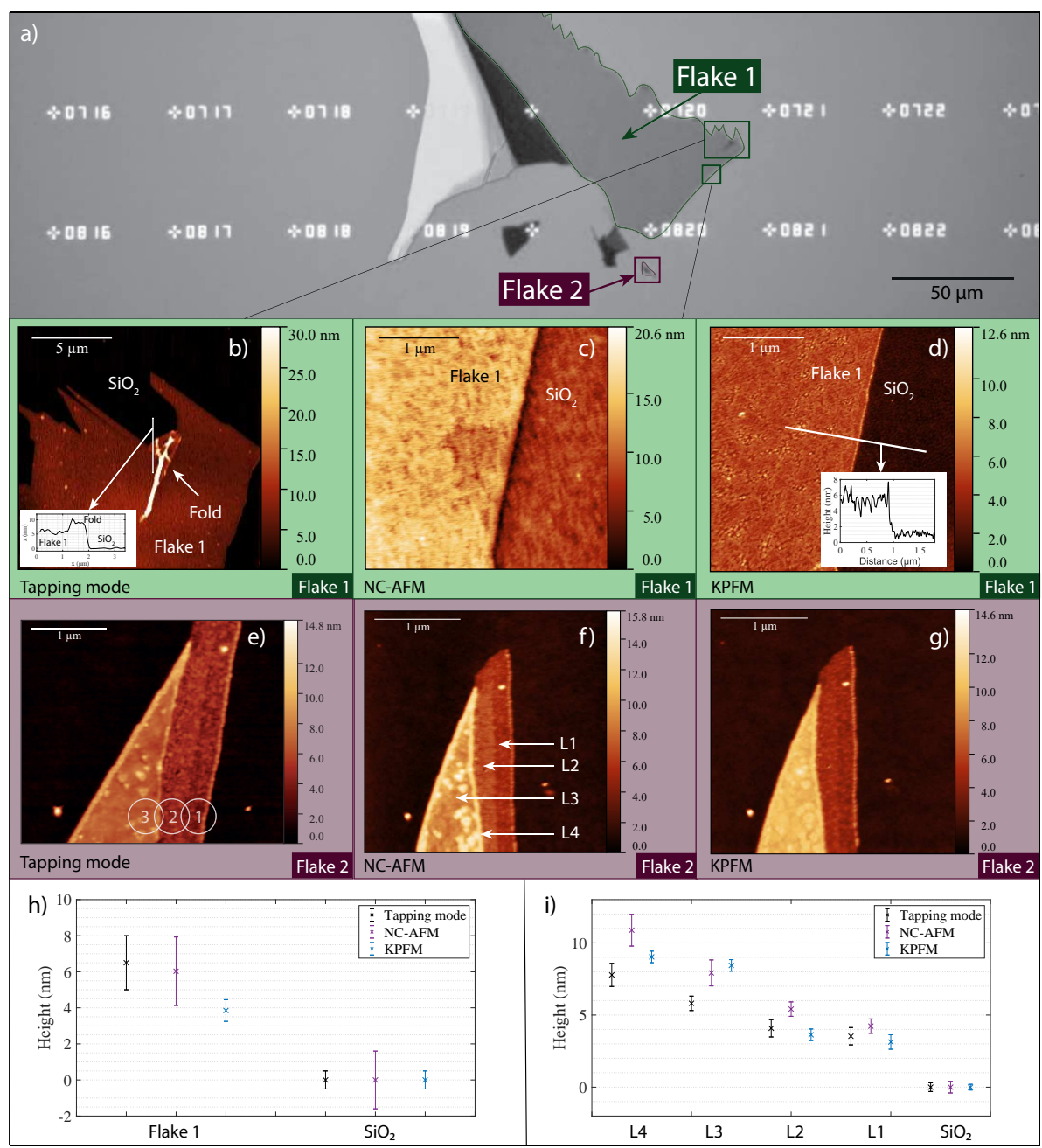}
\caption{(a) Optical microscopy image (red channel) of the MoSe$_2$ sample. Tapping mode, nc-AFM, and KPFM images of Flake 1 (b-d) and Flake 2 (e-g). (h-i) show the height measurements extracted from the AFM images shown. The height measurements were extracted from the images by averaging the height over large areas in each layer, and comparing to the substrate background. The details of the analysis are given in Supplementary Materials.}
\label{figureAFM}
\end{figure}

MoSe$_{2}$ flakes were transferred onto a substrate of SiO$_{2}$ using all-dry viscoelastic stamping \cite{Castellanos-Gomez2014DeterministicStamping}. Figure~\ref{figureAFM} provides an overview of the sample regions that were selected for study. Flake 1 is the largest, most visibly transparent area of the sample, and Flake 2 is a small island that has stepped layers. Figure~\ref{figureAFM}a shows a grayscale image of the red channel of the R, G, B channels of the color optical image (the red channel is selected here as it shows the highest degree of contrast between the layers). Figures~\ref{figureAFM}b-d and \ref{figureAFM}e-g show tapping mode, nc-AFM, and KPFM z-channel images of Flake 1 and Flake 2 respectively. The data in the KPFM z-channel images is the z-height measured after compensating for the contribution of the net electrostatic force to the height data. 

In the nc-AFM scan in Figure~\ref{figureAFM}f, Flake 2 appears to be stepped, with an overlayer of some additional features on the top layer. The apparent stepped layers and overlayer have been labeled L1, L2, L3, and L4 (note that these regions do not necessarily correspond to 1, 2, 3, and 4 atomic layers).  Figures~\ref{figureAFM}h-i show the height measurements extracted from the AFM images, which are presented in tabular form in the Supplementary Materials.

Comparing the optical contrast difference between Flake 1 and the substrate (values given for the color, green and blue channels in Supplementary Materials) for both the color image and red channel image to observations of other TMDCs (MoS$_2$ and WSe$_2$) on 300$\:$nm SiO$_2$/Si \cite{Li2013RapidMicroscopy}, the contrast of Flake 1 agrees with that of a monolayer rather than two stacked layers. However, a definite determination of the layer count based on optical contrast is difficult for a sample that is not consisting of large flakes of 1,$\:$2,$\:$3,$\:$...,$\:$N layers, where the transmittances from each flake can be compared relative to each other \cite{Taghavi2019ThicknessMaterials}. L.~Ottaviano et al \cite{Ottaviano2017MechanicalRevisited} demonstrated for MoS$_2$, that optical microscopy data alone can be misleading, as the contrast depends on the exact thickness of the SiO$_2$ layer on Si/SiO$_2$ samples, and is not always a monotonic function of the layer number \cite{Ottaviano2017MechanicalRevisited}. Optical microscopy could not be used to reliably characterize Flake 2 because it has dimensions smaller than the diffraction limit of light - another limitation of determining layer number by optical contrast \cite{Ottaviano2017MechanicalRevisited}.

\subsection{AFM}
The apparent AFM heights, shown in Figures~\ref{figureAFM}h-i, were measured by averaging large areas of the AFM images shown in Figures~\ref{figureAFM}b-g. Area averaging provides more robust, quantitative height measurements than individual line scans. See the Supplementary Materials for more information as well as the results of Figure~\ref{figureAFM}h-i presented in tabular form. Figures~\ref{figureAFM}h-i show that the height measurements of each AFM mode - tapping, nc-AFM and KPFM - often do not agree within uncertainty. This is because each mode provides different information about the sample properties, such as mechanical rigidity and electrostatic nonuniformity. For example, the step height between the substrate and Flake 1 in the KPFM scan (3.9$\:$$\pm$$\:$0.8$\:$nm) is significantly smaller than in the tapping mode and nc-AFM scans, suggesting that electrostatic forces inflate the height measurements when they are not compensated for. Similarly, the systematically lower heights measured with tapping mode on Flake 2 could be explained by mechanical compression of an overlayer or underlayer, since tapping mode is operated in a larger tip-sample force regime than nc-AFM and KPFM, so it is much more likely to mechanically influence (e.g. compress) the sample.

For each AFM mode, the substrate:Flake 1 and substrate:L1 heights measured with each AFM technique are all substantially higher than the $\sim\:0.7\:$nm expected for a single layer of MoSe$_2$. This result is similar to observations by L. Ottaviano et al.\cite{Ottaviano2017MechanicalRevisited}. Notably, the surface roughness of the sample measured with each AFM mode is also much larger than the atomic-scale roughness expected for clean MoSe$_2$ flakes. The line scan in the inset of Figure~\ref{figureAFM}d shows an example of the surface roughness: On the SiO$_2$ substrate, the roughness is sub-nanometer, whereas on Flake 1 the roughness is greater than 2~nm. Roughness characterizations of each AFM mode, shown in the Supplementary Materials, yield similar results. This suggests that the sample has a rough overlayer or underlayer, or is highly defective. Such a layer may have been left behind during sample fabrication. 

At the location where Flake 1 is folded, shown in the tapping mode data in Figure \ref{figureAFM}b, the step height between Flake 1 and the fold is smaller than the step between the SiO$_2$ substrate and Flake 1, as indicated by the line trace in the inset. Note that the height of the fold could not be accurately determined using the area averaging method, since the fold is too small. The height measurements from the line trace only are shown in Supplementary Materials. However, this difference in step height could further indicate the presence of an underlayer between the bottom layer and the substrate which is not present between stacked layers. 

The most notable electrostatic feature is L4 on Flake 2, shown in Figure~\ref{figureAFM}f-g: With nc-AFM, L4 is several nanometers high, whereas with KPFM compensation L4 is indistinguishable within the surface roughness from L3. The features comprising L4, which could be either on top of or underneath Flake 2, remained unchanged in shape and location even after the sample was annealed several times at 130~$^\circ$C in UHV for eight hours. L4 could be surface contamination introduced during the sample fabrication procedure: Selenium-based TMDCs that have been mechanically exfoliated under ambient conditions have been shown to be highly susceptible to airborne contaminants which are mobile on the surface and aggregate to larger patches with average height of 2.2~nm over 45 hours, influencing the apparent monolayer height and interlayer spacing\cite{Purckhauer2019AnalysisConditions}. 

Another notable feature shown in Figures~\ref{figureAFM}f-g is that the apparent step height between L1 and L2 in the nc-AFM image becomes zero within error when electrostatic forces are compensated with KPFM. E. Pollmann et al. showed for MoS$_2$ that the degree of screening of charge transfer to the SiO$_2$ substrate can depend on variation in water layers lying underneath the surface. The large uncertainty due to the roughness prevents a clear determination of whether an atomic step is present, but this result shows that non-uniform charging alone (e.g. due to variation in underlying water layers) can also simulate the presence of a step, and must be compensated for.

\subsection{Photoluminescence data}
\begin{figure}[ht]
\centering
\includegraphics[width=\textwidth]{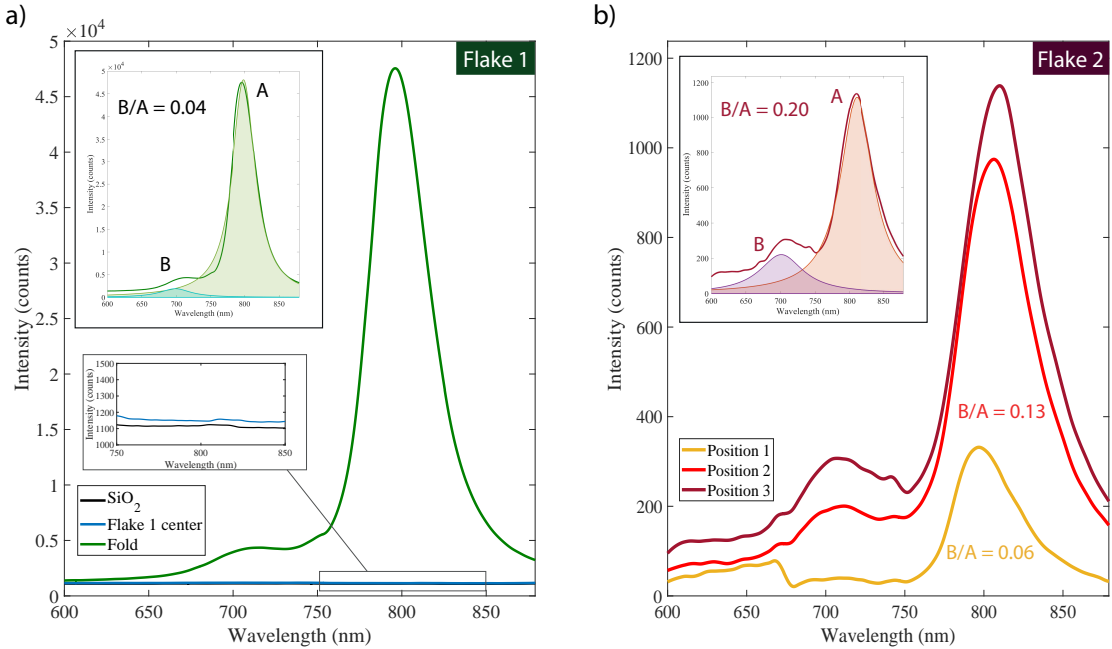}
\caption{PL spectroscopy data from Flake 1 (a) and Flake 2 (b) with insets showing peak fits from the sum of two Lorentzian functions to extract the B/A ratio. The full details of the fittings can be found in Supplementary Materials. The blue line in (a) shows there is no peak detected at the center of Flake 1, however a strong peak is observed at (798~$\pm$~2)~nm where the flake has a fold (black line). The PL spectra from Flake 2 in (b) were taken at the locations indicated in Figure \ref{figureAFM}e. Flake 2 showed main peaks at (801~$\pm$~3)~nm~(i), (808~$\pm$~3)~nm~(ii) and (810~$\pm$~4)~nm~(iii), with a shift towards a higher intensity but lower energy as the number of layers increases from right to left.  }
\label{figurePLdata}
\end{figure}

The results of characterization using PL spectroscopy of Flake 1 and Flake 2 are shown in Figure~\ref{figurePLdata}. The signal measured at the center of Flake 1 (Figure~\ref{figurePLdata}a, blue) is no greater than the SiO$_2$ background signal. This could be due to the presence of a water film between the SiO$_2$ and MoSe$_2$; Similar PL signal quenching has been observed for exfoliated MoS$_2$ monolayers, and it was determined to be caused by intercalated water between the substrate and the MoS$_2$ \cite{Pollmann2020ApparentExfoliation,Varghese2015TheDisulfide}. Previously reported weak PL intensities\cite{Jadczak2017ProbingExperiments,Tongay2012ThermallyMoS2}, or peak broadening have also been thought to be caused by aging effects such as oxidation and contamination of adsorbates \cite{Gao2016AgingMonolayers}. A high intensity peak (Figure~\ref{figurePLdata}a, green) was detected at the corner of Flake 1 where a fold is visible in Figures~\ref{figureAFM}a and b. The PL peak from this fold lies at a wavelength of 798$\:$$\pm$$\:$2$\:$nm. This agrees with the predicted position, peak shape and linewidth that have been reported for both a monolayer\cite{Tonndorf2013PhotoluminescenceWSe_2,Tongay2012ThermallyMoS2} and a bilayer data\cite{Sha2017Layer-by-layerEtching}. Depending on the degree of coupling with the underlying layer \cite{Froehlicher2018ChargeHeterostructures,Tongay2012ThermallyMoS2}, the fold could be behaving as a suspended monolayer with contributions from a bilayer signal.

The dominant peak at 800~nm is the A-exciton (ground state exciton) contribution\cite{McCreary2018A-Monolayers,Lu2014Large-areaSubstrates}, and a second smaller peak is present at $\sim$~91~nm below (i.e. at $\sim$~200~meV higher energy), which corresponds to the B-exciton (higher spin-orbit split state). The ratio of the intensities of the A and B peak is linked to the number of defects in the sample, where a higher B/A ratio indicates higher defect density \cite{McCreary2018A-Monolayers}. Fitting the data from the fold in Figure~\ref{figurePLdata}a with the sum of two Lorentzian peaks, we obtain the results shown in the inset. The B/A ratio is 0.04, suggesting a relatively high defect density compared to the values measured by K. McCreary et al. \cite{McCreary2018A-Monolayers}, where the highest B/A ratio observed for MoSe$_2$ monolayers synthesized by CVD was 0.025.

The PL spectra acquired at three positions on Flake 2 are shown in Figure~\ref{figurePLdata}b. The approximate position of the center of the laser beam corresponding to these three positions are labelled in Figure~\ref{figureAFM}e. Since the laser beam area was larger than the individual steps in the flake, the measured peaks are expected to be a convolution of contributions from the different layers. The spectrum at position 1 (yellow) has the A peak at (801~$\pm$~3)~nm, in agreement with the peak position of the fold in Figure~\ref{figurePLdata}a, and hence of a monolayer as observed in \cite{Tonndorf2013PhotoluminescenceWSe_2,Tongay2012ThermallyMoS2}. As the spot moves over the island from right to left, the peak position shifts towards 810 nm, as expected for an increasing number of layers \cite{Tonndorf2013PhotoluminescenceWSe_2,Wang2019SynthesisDeposition,Sha2017Layer-by-layerEtching}. The B/A ratio increases as the number of layers increases from right to left, and the top layer has a very high ratio of 0.20. Here, we do not see the intensity decreasing as the number of layers increases. This could be due to decoupling of the layers \cite{Arora2015ExcitonLimit}, but is also likely influenced by the fact that the laser beam spot is large and hence also partly covers the SiO$_2$ in for example the measurement at position 1, and hence sees a smaller contribution from MoSe$_2$. The results demonstrate how the presence of water and/or defects that are not visible optically can affect the PL peaks, highlighting the importance of thorough spatial characterization at the nm-scale in order to uncover the quality of the sample.

\section{Conclusion}

MoSe$_2$ flakes transferred onto SiO$_2$ were characterized using optical microscopy, PL spectroscopy and three modes of AFM. We showed complete quenching of the photoluminescence peak on the large Flake 1, which may be caused by an underlayer of water or other contamination. Electrostatic nonhomogeneities, a large uniform surface roughness and mechanical compressibility of the sample all indicate the presence of an over- or underlayer, which would also explain higher-than-expected layer heights measured in AFM. We have shown that the height of MoSe$_2$ layers on SiO$_2$ depends on the condition of the sample, and that it is vital to interpret step heights measured using AFM methods correctly. 

The approach typically used to count the number of layers in a TMDC sample is to combine an optical measurement with AFM data. Here we have demonstrated that great care must be taken when using these methods, as the layer count can be difficult to deconvolute from effects of charging, defects and contaminants, and any one method alone can lead to an incorrect or incomplete characterization of the sample. 

\section{Methods}
MoSe$_2$ layers were exfoliated onto SiO$_2$ using all-dry viscoelastic stamping. This process uses mechanical exfoliation to transfer 2D crystals onto a viscoelastic Gelfilm stamp, which in turn is pressed against the desired substrate to transfer the flakes onto the surface \cite{Castellanos-Gomez2014DeterministicStamping}. Optical microscopy images were obtained using a Nikon optical microscope (Nikon Eclipse LV150N) in reflective mode with a 50x times magnification objective. 

Tapping mode measurements were performed in air with a MFP3D (BIO) Asylum microscopy using 240C-PP OPUS tips with 1-2~N/m spring constant and oscillation amplitudes between 14-16~nm. Nc-AFM and KPFM measuremens were performed in ultrahigh vacuum after having been annealed at 130$^\circ$C for 8 hours with a modified JEOL JSPM-4500A UHV surface science system at room temperature, using Nanosensors platinum-iridium coated silicon tips with resonant frequency 330~kHz and spring constant 42~N/m. 

Photoluminescence spectra were collected at room temperature using a 532~nm laser excitation with a power of 150~µW and 30~s exposure time. A 100X objective was used to focus the laser beam on the sample. The spot size diameter was approximately 1 µm.

\section*{Acknowledgments}
The authors thank Philipp Nagler for the fabrication of the MoSe$_2$ sample and the acquisition of the optical microscopy image at Universität Regensburg. Funding from NSERC and FRQ-NT are gratefully acknowledged.

\section*{References}

\bibliographystyle{unsrt}

\bibliography{MoSe2references_manual}


\setcounter{figure}{0}
\renewcommand{\thefigure}{S\arabic{figure}}

\newpage
\section*{\Large Supplementary Materials} 

\newpage
\section*{Optical microscopy contrast measurements}

\begin{figure}[h]
\centering
\includegraphics[width=1\textwidth]{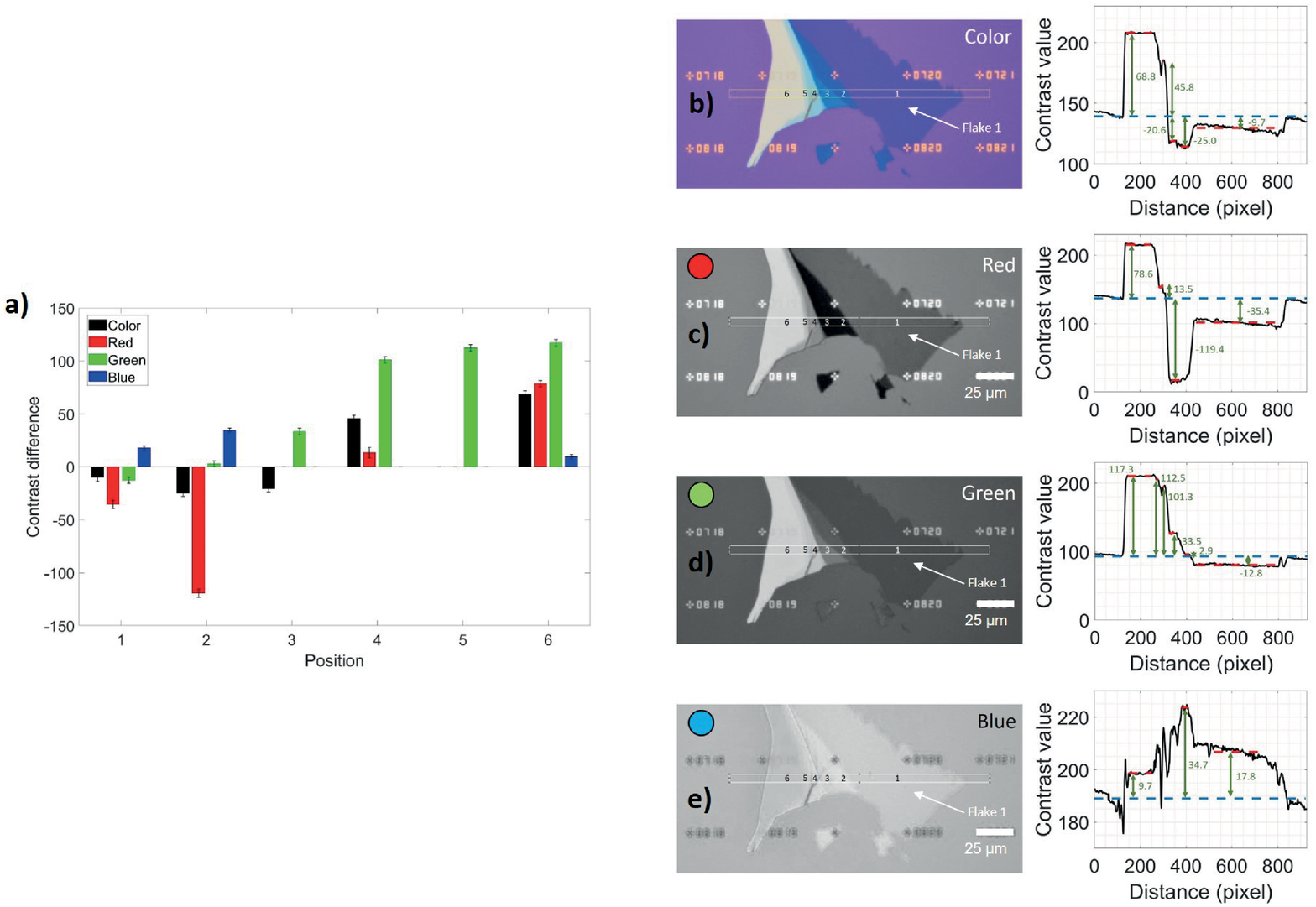}
\caption{Optical microscopy images and contrast profiles. The contrast difference between each position along the region shown by the white outline in b)-e) and the substrate were calculated following the procedure in \cite{Li2013RapidMicroscopy} and summarised in a). Figures b)-e) show the color, red, green and blue channels of the optical microscopy image, with the corresponding intensity line profile from the indicated region. In each contrast profile graph, the substrate contrast is labelled with a blue dashed line. The contrast value at each position was obtained by averaging the data at locations labelled by the red dashed lines.}
\label{figureSuppOptical}
\end{figure}

From the optical microscopy image of the sample, the contrast difference between the substrate and the MoSe$_2$ flakes can indicate the number of layers stacked \cite{Li2013RapidMicroscopy}. Figure \ref{figureSuppOptical} shows the contrast data for a selected region of the sample that includes Flake 1. The images were analysed according to the method in \cite{Li2013RapidMicroscopy} using ImageJ and MATLAB. The contrast values at location 1 (Flake 1) from the color and red channels (values -9.7 and -35.4) are similar to the contrast values observed for monolayers of MoS$_2$ (Fig. 5 in \cite{Li2013RapidMicroscopy}) and WSe$_2$ (Fig. 7 in \cite{Li2013RapidMicroscopy}) on 300 nm of SiO$_2$/Si in \cite{Li2013RapidMicroscopy}. 

\newpage
\section*{AFM height measurements}

\begin{figure}[h]
\centering
\includegraphics[width=0.9\textwidth]{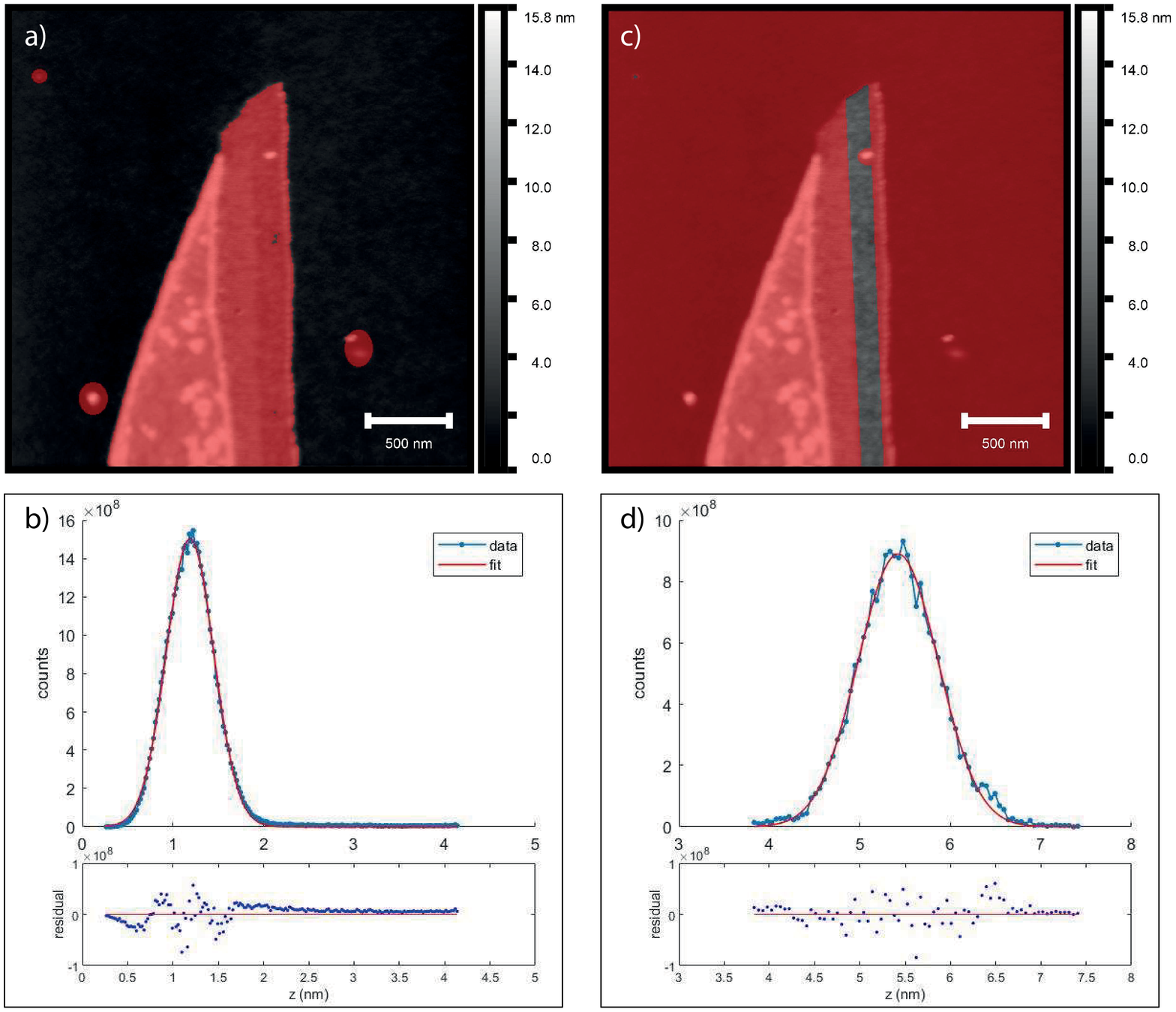}
\caption{An example of the procedure used to extract heights from AFM data. The total substrate area was masked (a), binned, and fit (b), as was one region of the sample (c-d). The fit means were then subtracted to give the sample region height.} 
\label{fig:Histograms}
\end{figure}

For all of the AFM data shown in this work, height measurements were found by averaging over large sample areas. This, as opposed to measuring individual line profiles, is a more robust way of determining the height, particularly if the sample surface is rough. First, the z-channel image is leveled until the width of a histogram of the entire substrate area is equal to the measurement noise. Then, Gaussean fits (of the form $f=a\times exp\left(\frac{-(z-b)}{c}\right)^2$) are found for each sample area. $b$ gives the mean z-channel value and $\sigma=\frac{c}{\sqrt{2}}$ gives the uncertainty of the z-channel value. The substrate mean was subtracted from each sample area mean to determine the sample area height ($h = b_{sample}-b_{substrate}$). Uncertainties were found by adding the Gaussean fit widths in quadrature as $\delta h = \sqrt{(\sigma_{sample})^2+(\sigma_{substrate})^2}$. Figure~\ref{fig:Histograms} shows an example of the masks and histograms used to evaluate the first layer height measured using ncAFM, where Figure~\ref{fig:Histograms}a and b show the substrate mask, histogram, and fit, and c and d show the first layer mask, histogram, and fit. The AFM height results are shown in Figure~1(h-i), and a summary is presented in tabular form in Table~\ref{table:Heights}.

\definecolor{light-gray}{gray}{0.8}
\newcolumntype{C}[1]{>{\centering\let\newline\\\arraybackslash\hspace{0pt}}m{#1}}
\vspace{0.5cm}
\begin{table}[h]
\centering
\begin{tabular}{|p{3cm}|p{4cm}|C{4cm}|}

     \hline
     Location & Method & Height (nm) \\
     \hline
    \rowcolor{light-gray}
    \multicolumn{3}{|l|}{Flake 1}\\
     \hline
    Substrate &  Tapping Mode & 0.0~$\pm$~0.5\\
      & NC-AFM & 0.0~$\pm$~1.6\\
      & KPFM & 0.0~$\pm$~0.5\\
      \hline
    1\textsuperscript{st} layer &  Tapping Mode & 6.5~$\pm$~1.5\\
      & NC-AFM & 6.0~$\pm$~1.9\\
      & KPFM & 3.9~$\pm$~0.6\\
      \hline

    \rowcolor{light-gray}
    \multicolumn{3}{|l|}{Flake 2}\\
     \hline
    Substrate &  Tapping Mode & 0.0~$\pm$~0.3\\
      & NC-AFM & 0.0~$\pm$~0.4\\
      & KPFM & 0.0~$\pm$~0.2\\
      \hline
    1\textsuperscript{st} layer &  Tapping Mode & 3.5~$\pm$~0.6\\
      & NC-AFM & 4.2~$\pm$~0.5\\
      & KPFM & 3.1~$\pm$~0.5\\
      \hline
    2\textsuperscript{nd} layer &  Tapping Mode & 4.1~$\pm$~0.6\\
      & NC-AFM & 5.4~$\pm$~0.5\\
      & KPFM & 3.6~$\pm$~0.4\\
      \hline
    3\textsuperscript{rd} layer &  Tapping Mode & 5.8~$\pm$~0.5\\
      & NC-AFM & 7.9~$\pm$~0.9\\
      & KPFM & 8.4~$\pm$~0.4\\
      \hline
    4\textsuperscript{th} layer &  Tapping Mode & 7.8~$\pm$~0.8\\
      & NC-AFM & 10.9~$\pm$~1.1\\
      & KPFM & 9.0~$\pm$~0.4\\
      \hline
\end{tabular}
\caption{Summary of the MoSe\textsubscript{2} heights measured with tapping mode AFM, NC-AFM, and KPFM for each region defined in Figure~1.}
\label{table:Heights}
\end{table}

\section*{\newpage AFM surface roughness measurements}
\definecolor{light-gray}{gray}{0.8}
\newcolumntype{C}[1]{>{\centering\let\newline\\\arraybackslash\hspace{0pt}}m{#1}}

The surface roughness was measured using Gwyddion by masking the various regions of the sample (as in Figure~\ref{fig:Histograms}) and measuring the root mean square (RMS) roughness. (Note that these roughness values very closely agree with the height uncertainties found by taking the histogram width of each layer, shown in Table~\ref{table:Roughnesses}.)

\vspace{0.5cm}
\begin{table}[h]
\centering
\begin{tabular}{|p{3cm}|p{4cm}|C{4cm}|}

     \hline
     Location & Method & RMS Roughness (nm) \\
     \hline
    \rowcolor{light-gray}
    \multicolumn{3}{|l|}{Flake 1}\\
     \hline
    Substrate &  Tapping Mode & 0.4\\
      & NC-AFM & 1.1\\
      & KPFM & 0.4\\
      \hline
    1\textsuperscript{st} layer &  Tapping Mode & 1.7\\
      & NC-AFM & 1.1\\
      & KPFM & 0.7\\
      \hline

    \rowcolor{light-gray}
    \multicolumn{3}{|l|}{Flake 2}\\
     \hline
    Substrate &  Tapping Mode & 0.2\\
      & NC-AFM & 0.3\\
      & KPFM & 0.2\\
      \hline
    1\textsuperscript{st} layer &  Tapping Mode & 0.6\\
      & NC-AFM & 0.5\\
      & KPFM & 0.4\\
      \hline
    2\textsuperscript{nd} layer &  Tapping Mode & 0.6\\
      & NC-AFM & 0.6\\
      & KPFM & 0.5\\
      \hline
    3\textsuperscript{rd} layer &  Tapping Mode & 0.4\\
      & NC-AFM & 0.6\\
      & KPFM & 0.3\\
      \hline
    4\textsuperscript{th} layer &  Tapping Mode & 0.8\\
      & NC-AFM & 1.2\\
      & KPFM & 0.4\\
      \hline
\end{tabular}
\caption{RMS roughnesses measured with tapping mode AFM, NC-AFM, and KPFM for each region defined in Figure~1.}
\label{table:Roughnesses}
\end{table}

\newpage
\section*{Flake 1 fold line trace}
A line trace taken across the fold on Flake 1 is shown in Figure \ref{fig:Flake1_fold_trace} below, with height measurements taken as the average of the z height along the single line. The uncertainty on the height value is given by the standard deviation of the z data along each line indicated in the graph. The small size and irregularity of the fold prevents a height measurement by the area averaging method, and the purpose of the line trace shown is to demonstrate a comparison of the step heights rather than accurate height values.

\begin{figure}[h]
\centering
\includegraphics[width=1\textwidth]{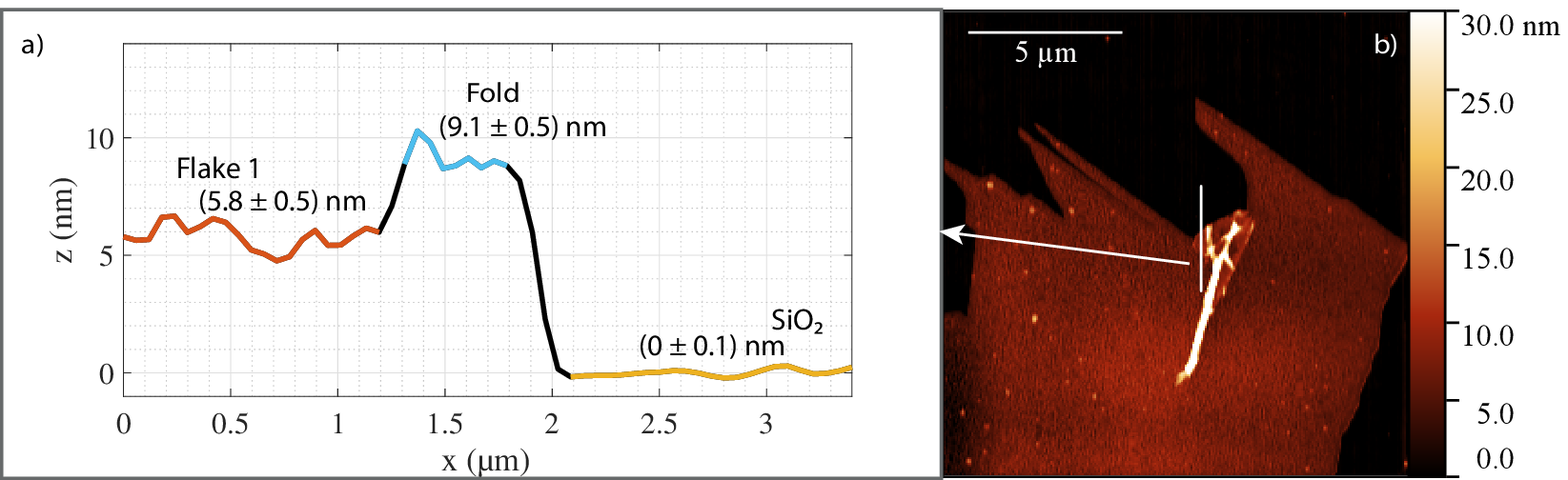}
\caption{Line trace from the fold on Flake 1. (a) shows the height measurements from the line trace only, with uncertainties given by the standard deviation of the z data at each location. Note that this method is not accurate compared to the area averaging method described under 'AFM height measurements' and hence cannot give a reliable quantitative result. Despite this, the line trace demonstrates clearly that the step height from the SiO$_2$ substrate to the first layer is higher than the step between the first layer and the fold. (b) tapping mode image of Flake 1 with the location of the line trace across the fold indicated.} 
\label{fig:Flake1_fold_trace}
\end{figure}

\newpage
\section*{Photoluminescence measurements}
The photoluminescence spectra in Figure 2 were analyzed and plotted in MATLAB. To remove random background noise artefacts in the data, the spectra were cleaned up using the function smooth(data,0.05,'rloess'), a local regression using weighted linear least squares and a 2$\mathrm{^{nd}}$ degree polynomial model. Lower weight is assigned to outliers in the regression; the method assigns zero weight to data outside six mean absolute deviations. The data was smoothed using this method using a span of 5$\%$ of the total number of data points. The background signal from the SiO$_2$ was subtracted from the MoSe$_2$ curves. The peak position and associated uncertainty were obtained as follows: The main peaks were fitted to a Lorentzian function $f(x) = A / (1 +((x-x_0)/ \gamma )^2),$ 
where $A$ is the height of the peak, $x_0$ is the position of the peak (wavelength), and $\gamma$ is the width of the peak at half maximum \cite{McCreary2018A-Monolayers}. The fitting was carried out in MATLAB (using fit function), with the appropriate starting parameters estimated. The error in the $x_0$ peak locations was found by forcibly varying $x_0$ around the fitted output value (while letting $A$ and $\gamma$ vary), to find the range of $x_0$ for which the coefficient of determination (r-squared) value of the fit stayed within 0.01 of its optimum. R-squared indicates the proportionate amount of variation in the response variable y explained by the independent variables x in the linear regression model. The larger the r-squared is, the more variability is explained by the linear regression model. R-squared is the proportion of the total sum of squares explained by the model: $ R_2 = SSR/SST = 1 - SSE/SST $ where $SSR$ is the sum of squared regression, $SST$ is the sum of squared total, and $SSE$ is the sum of squared error. 

\begin{figure}[h]
\centering
\includegraphics[width=0.65\textwidth]{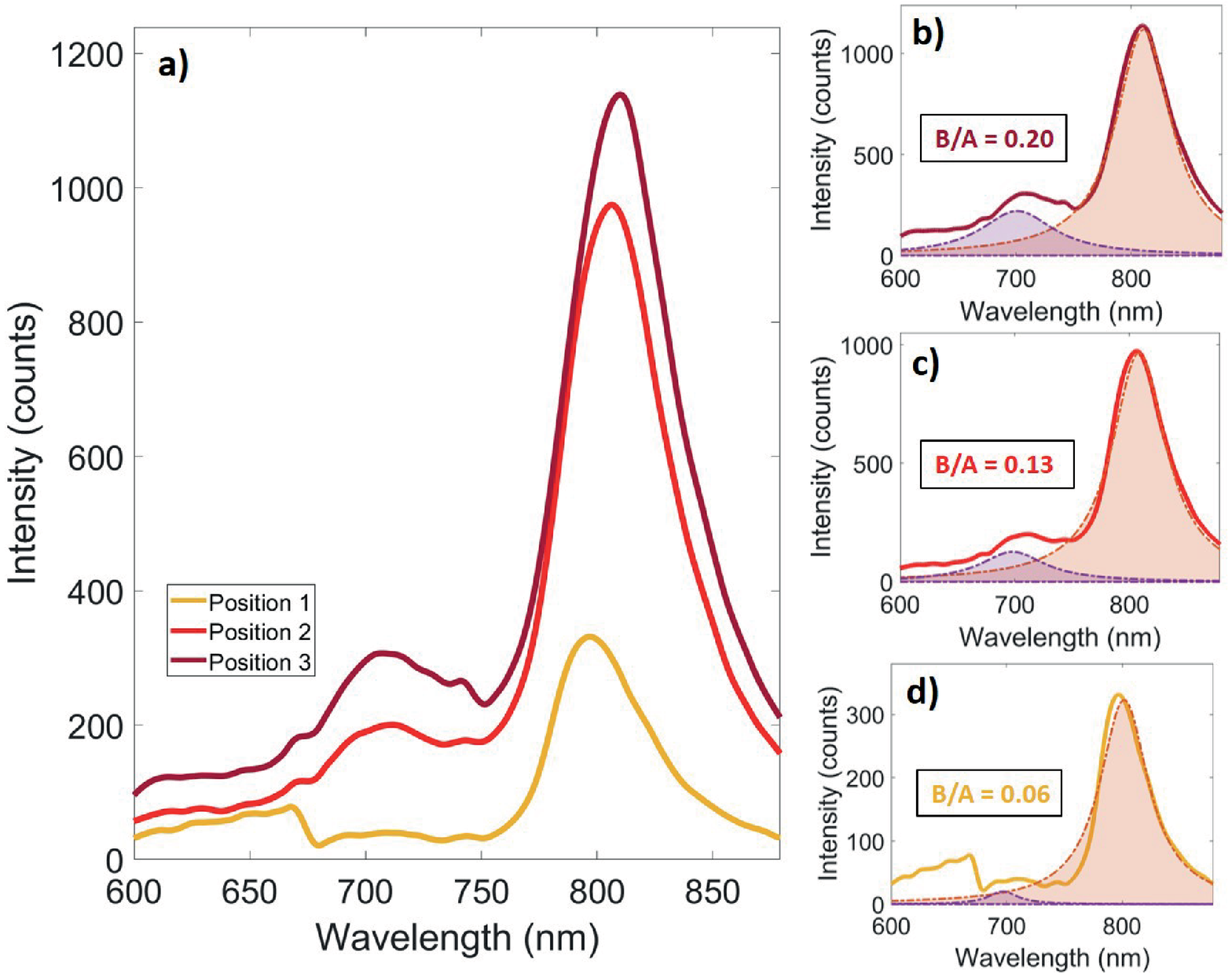}
\caption{Two peak Lorentzian curve fittings for positions 1, 2 and 3 on Flake 2. The ratio in intensity between the B and A peak is given for each spectrum.} 
\label{fig:Supp PL}
\end{figure}

\end{document}